\documentclass[fleqn,10pt]{wlscirep}
\title{Digital-Analog Quantum Simulation of Spin Models in Trapped Ions}

\author[1,*]{I\~nigo Arrazola}
\author[1]{Julen S. Pedernales}
\author[1]{Lucas Lamata}
\author[1,2]{Enrique Solano}
\affil[1]{Department of Physical Chemistry, University of the Basque Country UPV/EHU, Apartado  644, 48080 Bilbao, Spain}
\affil[2]{IKERBASQUE, Basque Foundation for Science, Maria Diaz de Haro 3, 48013 Bilbao, Spain}

\affil[*]{iarrazola003@gmail.com}

\begin{abstract}
We propose a method to simulate spin models in trapped ions using a digital-analog approach, consisting in a suitable gate decomposition in terms of analog blocks and digital steps. In this way, we show that the quantum dynamics of an enhanced variety of spin models could be implemented with substantially less number of gates than a fully digital approach. Typically, analog blocks are built of multipartite dynamics providing the complexity of the simulated model, while the digital steps are local operations bringing versatility to it. Finally, we describe a possible experimental implementation in trapped-ion technologies.
\end{abstract}
\begin{document}

\flushbottom
\maketitle
%
%
\thispagestyle{empty}

\section*{Introduction}

Quantum simulators are devices designed to mimic the dynamics of physical models encoded in quantum systems, enjoying high controllability and a variety of accessible regimes~\cite{Feynman82}.  It was shown by Lloyd~\cite{Lloyd96} that the dynamics of any local Hamiltonian can be efficiently implemented in a universal digital quantum simulator, which employs a universal set of gates upon a register of qubits. Recent experimental demonstrations of this concept in systems like trapped ions~\cite{Lanyon11} or superconducting circuits~\cite{Salathe15, Barends15, Barends15a} promise a bright future to the field. However, the simulation of nontrivial dynamics requires a considerable number of gates, threatening the overall accuracy of the simulation when gate fidelities do not allow for quantum error correction. Analog quantum simulators represent an alternative approach that is not restricted to a register of qubits, and where the dynamics is not necessarily built upon gates~\cite{Buluta09, Georgescu14}. Instead, a map is constructed that transfers the model of interest to the engineered dynamics of the quantum simulator. An analog quantum simulator, unlike digital versions, depends continuously on time and may not enjoy quantum error correction. In principle, analog quantum simulators provide less flexibility due to their lack of universality. 

Trapped-ion technologies represent an excellent candidate for the implementation of both, digital and analog quantum simulators~\cite{Blatt12}. Using electromagnetic fields, a string of ions can be trapped such that their motional modes display bosonic degrees of freedom, and two electronic states of each atom serve as qubit systems. Currently, trapped-ion techniques offer one of the highest degrees of controllability among quantum technologies, with high fidelity single- and two-qubit gates, and high readout precision~\cite{Harty14, Ballance15}. A wide variety of proposals for either digital or analog quantum simulations exist~\cite{Casanova12,Mezzacapo12,Casanova11,Alvarez13,Hayes14,Cheng15,Leibfried02}, and several experiments have demonstrated the efficiency of these techniques in trapped ions, in the digital~\cite{Lanyon11}, and analog cases. Examples of the latter include the quantum simulation of spin systems~\cite{Porras04,Friedenauer08, Kim10,Bermudez11,Britton12,Islam13,Jurcevic14,Richerme14} and relativistic quantum physics~\cite{Lamata07,Gerritsma10,Gerritsma11}. 
 
In this article, we propose a merged approach to quantum simulation that combines digital and analog methods. We show that a sequence of analog blocks can be complemented with a sequence of digital steps to enhance the capabilities of the simulator. In this way, the larger complexity provided by analog simulations can be complemented with local operations providing flexibility to the simulated model. More precisely, we show that analog quantum simulations of a restricted number of spin models can be extended to more general cases, as the Heisenberg model, by the inclusion of single-qubit gates. Our proposal is exemplified and validated by numerical simulations with realistic trapped-ion dynamics. We have named our approach digital-analog quantum simulation (DAQS), a concept that may be cross-linked to other quantum technologies.

\begin{figure}[ht]
\centering
\includegraphics[width=\linewidth]{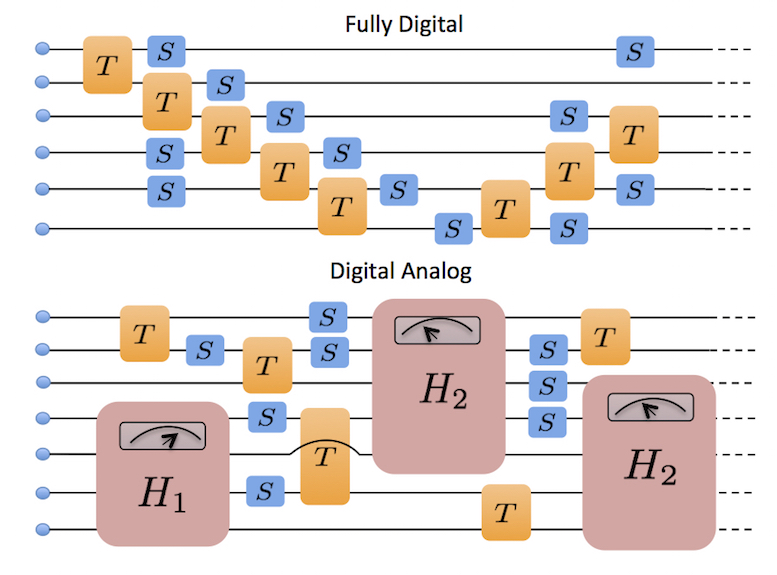}
\caption{\textbf{Fully Digital vs. Digital-Analog.} We depict the circuit representation of the digital and digital-analog approaches for quantum simulation. The fully digital approach is composed exclusively of single-qubit ($S$) and two-qubit ($T$) gates, while the digital-analog one significantly reduces the number of gates by including analog blocks. The latter, depicted in large boxes ($H_1$ and $H_2$), depend on tunable parameters, represented by an analog indicator, and constitute the analog quantum implementation of a given Hamiltonian dynamics.}\label{fig:Fig1}
\end{figure}

The proposed digital-analog quantum simulator is built out of two constitutive elements, namely, analog blocks and digital steps (see Fig.~\ref{fig:Fig1}). Digital steps consist of one- and two-qubit gates, the usual components of a universal digital quantum simulator. On the other hand, analog blocks consist in the implementation of a larger Hamiltonian dynamics, which typically involve more degrees of freedom than those involved in the digital steps. In general, analog blocks will depend on tunable parameters and will be continuous in time. 

\section*{Results}

To illustrate the digital-analog paradigm, we propose a nontrivial task: the quantum simulation of a generic spin-$\!1\!/\!2$ Heisenberg model in trapped ions. Its Hamiltonian reads ($\hbar=1$)
\begin{equation}\label{HB}
H_{\rm{H}}=\sum_{i<j}^N J_{i\!j} \ {\vec{\sigma}}_{\!i}\cdot{\vec{\sigma}}_{\!j}=\sum_{i<j}^N J_{i\!j} (\sigma_i^x \sigma_{\!j}^x+\sigma_i^y \sigma_{\!j}^y+\sigma_i^z \sigma_{\!j}^z),
\end{equation}
where the vector of Pauli matrices ${\vec{\sigma}}_i=(\sigma_i^x,\sigma_i^y,\sigma_i^z)$ characterizes the spin of particle $i$, while $J_{i\!j}$  is the coupling strength between spins $i$ and $j$. To simulate this model, we will consider off-the-shelf interactions of ion chains~\cite{Jurcevic14,Richerme14}. More specifically, spin Hamiltonians
\begin{eqnarray}
H_{XX}=\sum_{i<j}J_{i\!j} \sigma_i^x \sigma_{\!j}^x, \ \ 
H_{XY}=\sum_{i<j}J_{i\!j} (\sigma_i^x \sigma_{\!j}^x +\sigma_i^y \sigma_{\!j}^y) 
\label{analogblock}
\end{eqnarray}
can be used as analog blocks, while single-qubit rotations $R_{x,y}(\theta)= \exp({-i\theta\sum_i \sigma_i^{x,y}})$ perform digital steps.

\begin{figure}[ht]
\centering
\includegraphics[width=\linewidth]{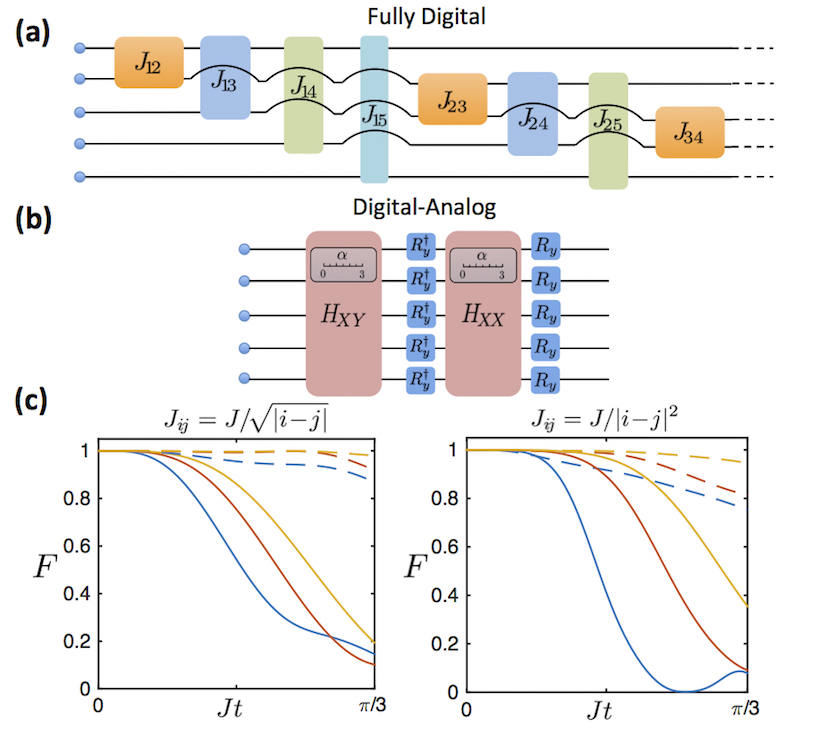}
\caption{\textbf{Digitization of the Heisenberg model.} (a) Scheme of a Trotter step of a purely digital quantum simulation for a generic spin dynamics with five sites. (b) Trotter step of a Digital-Analog protocol for the simulation of the Heisenberg model with tunable $\alpha$. (c) Fidelity loss obtained with the application of fully digital (solid lines) and digital-analog (dashed lines) protocols for the initial state $|\!\! \downarrow \downarrow \uparrow \downarrow \downarrow \rangle$. Blue (lower), orange (middle), and yellow (upper) colours represent one, two, and three Trotter steps, respectively. For the digital case, fidelity $F$ decays faster with $t$ for long-range interactions, while $F$ remains similar for the digital-analog protocol.}\label{fig:Fig2}
\end{figure}

It is known that if a Hamiltonian can be decomposed into a sum of local terms, $H=\sum_{k} H_k$, its dynamics $U=e^{-iHt}$ can be approximated by discrete stepwise unitaries, according to the Trotter formula
\begin{equation}\label{Trotter}
U=(\prod_{k} e^{-iH_kt/ l})^l + O(t^2/l),
\end{equation}
where $l$ is the number of Trotter steps.  Here, the error of the approximation to the second order $O(t^2/l)$ is bounded by $\|O(t^2/l)\|_{\rm{sup}} \leq \sum_{k=2}^{\infty} l\|Ht/l\|^k_{\rm{sup}}/k!$ Thus, the digital error will decrease for a larger number of Trotter steps $l$. For the specific case of the antiferromagnetic Heisenberg Hamiltonian ($J_{i\!j}>0$), we have that $\|H\|_{\rm{sup}}=\sum_{i<j}^N J_{i\!j}$. This indicates the growth of the digital error bound with the number of spins in the chain, $N$, and with the range of the interaction between the spins. On the other hand, each particular decomposition of the Hamiltonian will show a different truncation error, which will grow linearly with the sum of the commutators of all the Hamiltonian terms~\cite{Lloyd96}. For the Heisenberg Hamiltonian, a suitable decomposition is given by $H_{\rm{H}}=H_{XY}+H_{ZZ}$. The dynamics of the $H_{ZZ}=\sum_{i<j}J_{i\!j} \sigma_i^z \sigma_{\!j}^z $ term can be generated with the proposed DAQS protocol, by combining the global qubit rotation $R_y(\pi/4)$ with the Ising-like dynamics $H_{XX}$. In this case, a Trotter step is given by the decomposition
\begin{equation}\label{UnitaryEvolution}
U^{\rm{H}}(t/l)=e^{-i H_{XY}t/l} R_y e^{-i H_{XX}t/l}R_y^\dagger ,
\end{equation}
where $R_y \equiv R_y(\pi/4)$. In Fig.~\ref{fig:Fig2}a and \ref{fig:Fig2}b, we show the circuit representation of the simulation algorithm following such a Trotter decomposition, as compared to its equivalent in a purely digital quantum simulator, that is, a simulator built only upon one- and two-qubit gates. The latter will need to include in the algorithm a two-qubit gate for each two-body interaction contained in the Hamiltonian. Even though these elementary gates can be realized with high accuracy, one needs to apply a large number of them, especially when the model has a long interaction range. The induced global fidelity loss is given not only by the imperfection of experimental gates, but also by the noncommutativity of these gates, which increases the Trotterization error. The inclusion of analog blocks like $H_{XY}$ and $H_{XX}$, accessible in trapped ions~\cite{Jurcevic14,Richerme14}, can become beneficial for the simulation of many-qubit spin models.

In Fig.~\ref{fig:Fig2}c, we plot the fidelity of the time evolution of two particular coupling regimes of the Heisenberg Hamiltonian for different numbers of Trotter steps, and we compare them to the fidelity of a purely digital algorithm. Numerical results show that the digital-analog approach achieves higher fidelities at all studied times for both models. Furthermore, the DAQS method represents a higher advantage with respect to the digital approach when the interaction range of the simulated model is longer. In general, a long-range Hamiltonian has more noncommuting terms that contribute to a larger digital error. DAQS takes advantage of its versatility in the Hamiltonian decompositions, as given by the sum of only two terms in the considered example. These terms do not always commute, but the associated commutator happens to be small for long-range spin interactions. Actually, for the limiting $J_{i\!j}=J$ case, DAQS produces no digital error, i.e., the analog blocks commute. In consequence, we consider this approach to represent a solid alternative for simulating generic long-range Heisenberg models. 

As we already mentioned, the digital-analog protocol  shown in Fig.~\ref{fig:Fig2}b needs two analog blocks per Trotter step, independently of the number of spins $N$. On the contrary, the number of entangling gates in a fully digital protocol grows with $N$. For generating each two-body interaction, at least a two-qubit gate is needed, and the number of two-body interactions will vary depending on the simulated model. This ranges from $N-1$, in the case of nearest neighbour interactions, to $N(N-1)/2$, in the long-range interaction case. Apart from the Trotter error, any realistic digital simulation has to deal with errors arising from the imperfection of experimental gates, which we quantify by the gate infidelity. In this respect, and in the long-range case, DAQS leads to a better result as long as the analog-block gate infidelity fulfills $\epsilon_{AB}\leq \frac{N(N-1)}{4}\epsilon_{T}$, where $\epsilon_{T}$ is the two-qubit-gate infidelity. Consequently, a purely digital proposal would need to compensate the larger number of gates with better gate fidelities~\cite{Ballance15}. However, it is fair to assume that the two-qubit gate fidelity will decrease when we increase the number of ions in the trap~\cite{Monz11}. The gate fidelity of the analog block, on the other hand, will also decrease with the size of the system. 

\subsection*{Proposal for an experimental implementation}

The experimental implementation of the considered digital steps, which correspond to local spin rotations, is easily achieved in trapped ions through carrier transitions~\cite{Haffner08}. The spin-spin interactions, corresponding to the proposed analog blocks, were first suggested by Porras \& Cirac~\cite{Porras04}, and have been implemented in several experiments~\cite{Friedenauer08,Kim10, Jurcevic14}. To show their derivation, we first consider a set of $N$ two-level ions confined in a linear trap, coupled to the $2N$ radial modes of the string by a pair of non-copropagating monochromatic laser beams. These lasers are oriented orthogonally to the ion chain, with a $45$ degree angle with respect to the $x$ and $y$ radial directions. We work in an interaction picture with respect to the uncoupled Hamiltonian $H_0=\frac{\omega_0}{2} \sum_j \sigma_j^z +\sum_m \nu_m a_m^\dagger a_m $. Here, $\omega_0$ is the frequency of the electronic transition of the two-level ion, and $\nu_m$ the frequency of the transverse motional mode $m$ of the ion string, with  annihilation(creation) operator $a_m$($a_m^\dagger$). The interaction Hamiltonian for the system reads
\begin{equation}\label{IntHamil}
H^I = \sum_{j=1}^N \Omega_{\!j}\sigma_j^+  e^{-i(\epsilon t-\phi_L)} \sin\big[{\sum_{m=1}^{2N}\!\!\eta_{j\!,m}(a_m e^{-i\nu_m t} + a^\dag_m e^{i\nu_m t})\big]} + {\rm{H.c.}},
\end{equation}
where $\Omega_j$ is the Rabi frequency of the laser for the $j$th ion, $\epsilon=\omega_L-\omega_0$ is the detuning of the laser frequency with respect to the electronic transition, $\phi_L$ is the laser phase, and $\eta_{j,m}$ is the Lamb-Dicke parameter, which is proportional to the displacement of the $j$th ion in the $m$th collective mode~\cite{James98}.

To obtain the effective spin-spin interactions, the two pairs of laser beams are tuned off-resonantly to the red and blue sidebands of the $2N$ radial modes with symmetric detunings ${\epsilon_{\pm}=\pm(\nu_{\rm{COM}}+\Delta)}$, where $\Delta \ll \nu_{\rm{COM}}$. Here, $\Delta$ denotes the detuning of the laser with respect to the first blue sideband of the motional mode with highest frequency. This corresponds to the center-of-mass (COM) mode in the radial $x$-axis, in the case where this axis has the highest trapping frequency ($\omega_x > \omega_y$). The Lamb-Dicke regime, which corresponds to keeping only the linear term in the expansion of the sine in Eq.~\ref{IntHamil}, can be considered when $|\eta_{j,m}|\sqrt{\langle a^\dag_ma_m\rangle} \ll 1$. Moreover, we can also neglect fast oscillating terms under the so called vibrational rotating-wave approximation (RWA), which holds when $|\eta_{j\!,m}\Omega_{\!j}| \ll \nu_m$. All in all, the resulting Hamiltonian is given by
\begin{equation}\label{BicHamil}
H_{\rm{bic}}=\sum_{j=1}^{N} \sum_{m=1}^{2N}\Omega_{\!j}\eta_{j\!,m} \big(\sigma_j^++\sigma_j^-\big) \big(a_me^{i\Delta_m t}+a_m^\dagger e^{-i\Delta_m t} \big),
\end{equation}
where $\Delta_m=\Delta + (\nu_{\rm{COM}}-\nu_m) > \Delta$. If $|\eta_{j\!,m}\Omega_{\!j}| \ll \Delta_m$, we can perform the adiabatic elimination of the motional modes, which are only virtually excited. As a result, a second order effective Hamiltonian with only spin-spin interaction terms arises\begin{equation}\label{effectiveHamil}
H_{\rm{eff}}=\sum_{i<j}^N J_{i\!j} \sigma_i^x \sigma_{\!j}^x=H_{XX},
\end{equation}
where the spin-spin coupling is given by
\begin{equation}\label{effectiveCoupling}
J_{i\!j}=2\Omega_i \Omega_{\!j} \sum_{m=1}^{2N} \frac{\eta_{i\!,m}\eta_{j\!,m}}{\Delta_m}\approx \frac{J}{|i-j|^\alpha},
\end{equation}
with $J\equiv {\sum_i J_{i,i+\!1}}/{(N\!-\!1)} > 0$ and tunable $0 < \alpha < 3$, see Britton {\emph{et al.}}~\cite{Britton12}.
In Fig.~\ref{fig:Fig3}a, we plot the spin-spin coupling matrix obtained for five $^{40}$Ca$^+$ ions, with the values $\Delta=(2\pi)60$kHz for the detuning, ${\Omega=(2\pi)62}$kHz for the Rabi frequency, $\vec{\omega}=(2\pi)(2.65,2.63,0.65)$MHz for the trapping frequencies and $\lambda=729$nm for the laser wavelength. The coupling matrix approximately follows the power-law decay with $\alpha \approx 0.6$, which essentially can be tuned varying $\Delta$ and $\omega_z$. Here we have assumed $\Omega_j=\Omega$, which is safe for the five ion chain that we are considering~\cite{Richerme14}. If we were to consider longer chains one would need to have into account that the laser intensity profile has a Gaussian shape and therefore that the outermost ions may have smaller Rabi frequency than the central ones. This would result in a modification of the coupling scaling law. The $XY$ Hamiltonian can be generated introducing a slight asymmetry in the detuning of the bichromatic laser $\epsilon_{\pm}=\pm(\nu_{\rm{COM}}+\Delta)+\delta$, $\delta \ll \Delta$. This introduces in the spin ladder operators a time-dependent phase factor, $\sigma^+ \rightarrow \sigma^+e^{-i\delta t}$ and $\sigma^- \rightarrow \sigma^- e^{i\delta t}$, making several terms in the effective Hamiltonian negligible under the RWA. The effective Hamiltonian, then, reads
\begin{equation}\label{effectiveHamil}
H_{\rm{eff}}=\sum_{i<j}^N J_{i\!j} (\sigma_i^+ \sigma_{\!j}^- +\sigma_i^- \sigma_{\!j}^+) + \frac{\delta}{\Delta}\sum_{j=1}^N B_{\!j} \sigma_j^z \approx \frac{1}{2} H_{XY},
\end{equation}
where $B_{\!j}=\Delta\Omega_j^2 \sum_{m}{\!(\!\eta_{j,m}/\Delta_m)}^2 (a_m^\dagger a_m \! +\!1\!/2) $. In addition to the $XY$ interaction, terms proportional to $ a_m^\dagger a_m\sigma_{\!j}^z$ appear. However, the contribution of these terms is smaller than the spin-spin term by a factor of $\delta/\Delta$ and can be neglected in the case where a small number of phonons is excited ($\langle B_j \rangle \sim J_{i\!j}$). The $XY$ Hamiltonian can also be implemented using a single monochromatic laser field tuned off-resonantly to the first blue sidebands of the $2N$ modes. As for the bichromatic case, the vibrational modes are only virtually excited and this gives rise to an effective spin-spin Hamiltonian. Nevertheless, in this case the strength of the terms $a_m^\dagger a_m\sigma_{\!j}^z $ is of the same order of magnitude as that of the spin-spin coupling term. This makes this last approach more sensitive to the heating of the phononic degrees of freedom.

\subsection*{Numerical simulations}

\begin{figure}[ht]
\centering
\includegraphics[width=\linewidth]{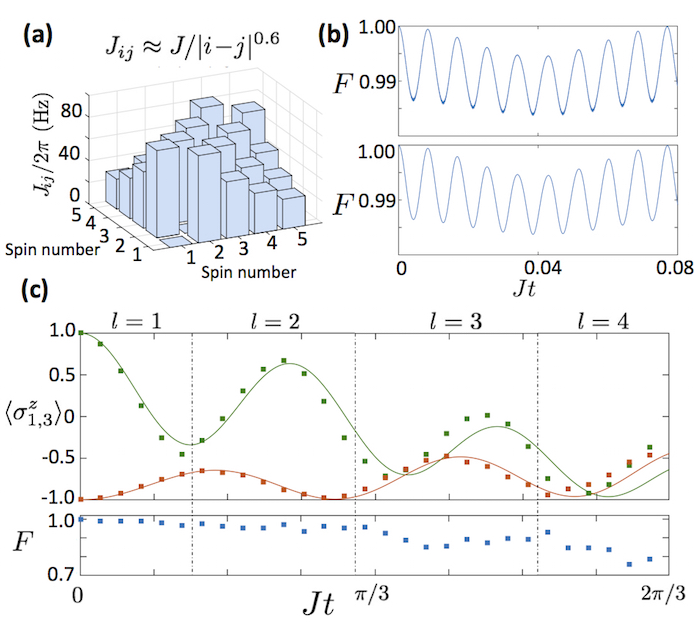}
\caption{ \textbf{Numerical simulations.} (a) Long-range ($\alpha \approx 0.6$) spin-spin coupling matrix $J_{ij}$ for $N=5$ spins.  (b) Fidelity of the $H_{XY}$ analog block for the state ${|\!\! \downarrow \downarrow \uparrow \downarrow \downarrow \rangle}$, with (lower plot) and without (upper plot) applying the vibrational RWA. The fidelity is periodic in time and thus we just plot one period. The numerical simulation assumes only the Lamb-Dicke regime, therefore accounting for the main sources of error that are the RWA and the adiabatic elimination. (c) Magnetization of the first (orange, lower curve) and third (green, upper curve) spins $\langle\sigma_{1,3}^z\rangle$ (upper plot) and the state fidelity of the digital-analog protocol (lower plot) versus time, for the protocol in Fig. \ref{fig:Fig2}b with initial state as in (b). Solid curves correspond to the ideal state produced by the Heisenberg Hamiltonian in Eq. (\ref{HB}), while dots correspond to the state produced by the DAQS approach. We divide the time interval into regions and simulate each time region using optimized numbers of Trotter steps, in order to maximize the state fidelity produced by our protocol.}\label{fig:Fig3}
\end{figure}

In Fig.~\ref{fig:Fig3}b, we depict the fidelity of an $H_{XY}$ analog block for five ions as a function of time. For numerical feasibility, instead of considering the ten radial modes present in the five ion case, we have considered a single COM mode with an effective Lamb-Dicke parameter $\eta^{\rm{eff}}\equiv \Omega^{-1}\!\sqrt{J \Delta /2}$ that represents the effect of all radial modes. This can be done as long as the chosen effective Lamb-Dicke parameter results in a coupling strength of the same order of magnitude of the one under study. This is true because the infidelity of the adiabatic approximation depends directly on the coupling strength J. Moreover, we choose the COM mode because it is the most unfavorable one for the approximation in terms of its detuning $\Delta$. In this manner we are able to give a safe fidelity estimate, overcoming the computationally demanding task of simulating the model with the ten motional modes. The analog blocks result from an effective second-order Hamiltonian, and their fidelity is subject to the degree of accuracy of the involved approximations. In the case of the $H_{XX}$ interaction, the greater the $\Delta$, the better the approximation and the gate fidelity. However, the simulation time is longer because $J_{i\!j}$ is inversely related to $\Delta$. The same is true for the $H_{XY}$ interaction, but the latter involves additional approximations that require $\delta \ll \Delta$ and $J \ll \delta$. For $\Delta=(2\pi)60$kHz and $\delta=(2\pi)3$kHz, the $H_{XY}$ gate infidelity can go up to $\epsilon_{AB}\approx 0.02$, as we can observe in Fig.~\ref{fig:Fig3}b. Obviously, the $H_{XX}$ analog block gives better results, since it is subject to fewer approximations.  We have also plotted the time evolution for the Hamiltonian in Eq. (\ref{BicHamil}), in which the vibrational RWA has been applied, and the Lamb-Dicke regime has been considered. It can be observed that there is no appreciable difference between both plots, which validates the vibrational RWA in the considered parameter regimes.

A numerical simulation of the dynamics produced by the digital-analog protocol in Fig.~\ref{fig:Fig2}b is presented in Fig. \ref{fig:Fig3}c. More precisely, we plot the magnetization of the first (orange, lower curve) and third (green, upper curve) spins, $\langle \sigma_1^z (t)\rangle$ and $\langle \sigma_3^z (t)\rangle$ (upper plot), and the fidelity associated with the digital-analog protocol (lower plot) as a function of time, in a five-ion chain. In order to maximize the fidelity of the quantum simulation, we need to reach a compromise between the number of Trotter steps, which increases the fidelity by reducing the digital error, and the total number of gates, which lowers the total fidelity by increasing the accumulated gate error. For that, we divide the time interval in regions, and we numerically simulate each region with the optimal number of Trotter steps. The fidelity of single-qubit gates is in general high~\cite{Harty14}, so we treat them as perfect in our calculation. As can be seen in Fig. \ref{fig:Fig3}c, we reach times of $Jt=2\pi/3$ with a state fidelity of approximately $70\%$, assuming $\Delta=(2\pi)60$kHz and $\delta=(2\pi)3$kHz. As we discussed, we could lower the error coming from the analog block by taking a larger value for $\Delta$ and, thus, improve the fidelity of the quantum simulation. However, this would increase the experimental time, which is limited by the coherence time of the system. For our analysis, we have considered real time dynamics of up to 13ms, which is below coherence times in trapped ion chains~\cite{Jurcevic14}.

\section*{Discussion}
We introduce the digital-analog approach to quantum simulations, which represents a solid alternative to universal digital quantum simulation, whenever gate fidelities are not high enough to allow for quantum error correction. We have shown that the DAQS approach is advantageous for the simulation of the Heisenberg model in trapped ions. Also, we have validated through numerical simulations that an implementation of our protocol is within experimental reach. With the proposed DAQS approach, we expect that a larger number of ions can be employed when compared with purely digital methods, reaching the size of analog quantum simulators~\cite{Jurcevic14,Richerme14}. The natural continuation of this research line is to explore how other models could benefit from the DAQS technique. Under the general argument that analog blocks concentrate the complexity of the model in high fidelity analog simulations, it is reasonable to expect that plenty of models will profit from such a simulation procedure. The central concepts of this novel approach are platform independent and, thus, can be exported to other quantum technologies. We consider the introduced DAQS techniques to be an important ingredient enhancing the toolbox of quantum simulations.

\section*{Acknowledgements}

We acknowledge support from a Basque Government PhD grant PRE-2015-1-0394,  a UPV/EHU PhD grant, Ram\'on y Cajal Grant RYC-2012-11391, UPV/EHU UFI 11/55, Spanish MINECO/FEDER FIS2015-69983-P,  and UPV/EHU Project EHUA14/04.

\section*{Author contributions statement}

I.A., J.S.P. and L.L. performed the main calculations and numerical simulations. I.A., J.S.P., L.L. and E.S. contributed to the generation and development of the ideas and to the writing of the paper.

\section*{Additional information}

\textbf{Competing financial interests:} The authors declare no competing financial interests.

\end{document}